\documentclass[%
twocolumn,
 amsmath,amssymb,
 aps,
]{revtex4-1}

\usepackage{graphicx}
\usepackage{multirow}
\usepackage{color}
\usepackage{dcolumn}
\usepackage{bm}

\begin{document}

\preprint{PRL/Weibel}

\title{Relativistic electron streaming instabilities modulate proton beams accelerated in laser-plasma interactions}

\author{S. G\"ode$^{1,2}$}
\author{C. R\"odel$^{1,3}$}
\author{K. Zeil$^4$}
\author{R. Mishra$^1$}
\author{M. Gauthier$^1$}
\author{F. Brack$^{4,6}$}
\author{T. Kluge$^4$}
\author{M. J. MacDonald$^{1,5}$}
\author{J. Metzkes$^4$}
\author{L. Obst$^{4,6}$}
\author{M. Rehwald$^{4,6}$}
\author{C. Ruyer$^{1,7}$}
\author{H.-P. Schlenvoigt$^4$}
\author{W. Schumaker$^1$}
\author{P. Sommer$^4$}
\author{T.E. Cowan$^{4,6}$}
\author{U. Schramm$^{4,6}$}
\author{S. Glenzer$^1$}
\author{F. Fiuza$^{1}$}
\email{fiuza@slac.stanford.edu}

\affiliation{$^1$High Energy Density Science Division, SLAC National Accelerator Laboratory, Menlo Park, California 94025, USA}
\affiliation{$^2$European XFEL GmbH, Holzkoppel 4, 22869 Schenefeld, Germany}
\affiliation{$^3$Friedrich-Schiller-University Jena, Max-Wien-Platz 1, 07743 Jena,Germany}
\affiliation{$^4$Helmholtz-Zentrum Dresden-Rossendorf, Institute of Radiation Physics, Bautzner Landstr. 400, 01328 Dresden, Germany}
\affiliation{$^5$University of Michigan, Ann Arbor, MI 48109, USA}
\affiliation{$^6$Technische Universit\"at Dresden, 01062 Dresden, Germany}
\affiliation{$^7$CEA, DAM, DIF, F-91297 Arpajon, France}

\begin{abstract}
We report experimental evidence that multi-MeV protons accelerated in relativistic laser-plasma interactions are modulated by strong filamentary electromagnetic fields. Modulations are observed when a preplasma is developed on the rear side of a $\mu$m-scale solid-density hydrogen target. Under such conditions, electromagnetic fields are amplified by the relativistic electron Weibel instability and are maximized at the critical density region of the target. The analysis of the spatial profile of the protons indicates the generation of $B>$10 MG and $E>$0.1 MV/$\mu$m fields with a $\mu$m-scale wavelength. These results are in good agreement with three-dimensional particle-in-cell simulations and analytical estimates, which further confirm that this process is dominant for different target materials provided that a preplasma is formed on the rear side with scale length $\gtrsim 0.13 \lambda_0 \sqrt{a_0}$. These findings impose important constraints on the preplasma levels required for high-quality proton acceleration for multi-purpose applications.
\end{abstract}

\maketitle
Proton acceleration from high-intensity laser-plasma interactions offers an important and compact tool for diagnosing high-energy-density matter and strong electromagnetic fields \cite{borghesi1,mackinnon,huntington} as well as for potential application to inertial fusion energy \cite{roth,fernandez} and medicine \cite{bulanov,zeil2013}. The control of the proton beam quality is critical for these multi-disciplinary applications. The majority of proton acceleration experiments have used $1 - 100 \,\mu$m thick solid-density foils as the main target \cite{snavely,fuchs,Zeil2010,macchi}. Protons are dominantly accelerated by the so-called Target Normal Sheath Acceleration (TNSA) mechanism \cite{wilks,mora}, where fast electrons produced by the laser at the front surface excite a strong space-charge electric field as they leave the target on the rear side. This E-field can reach amplitudes of several MV/$\mu$m and accelerate protons to 10s of MeV. Spatial modulations of the proton beams have been observed for insulator targets (\emph{e.g.} plastic, CH) \cite{roth2,fuchs2,manclossi,margarone2015} and attributed to an instability of the ionization front that affects the uniformity of the sheath E-field \cite{manclossi}.

More recently, a significant effort has been devoted to exploit the use of ultra-thin ($1 \mathrm{nm} - 1 \,\mu$m) and/or mass limited solid targets to control the energy and the quality of the proton beams \cite{Henig2009}. However, in this regime, experimental studies have also shown that accelerated proton beams can develop strong spatial modulations \cite{Palmer2012,Powell2015,Metzkes2014}. These modulations were attributed to Rayleigh-Taylor-type (RT) instabilities at the front side of the laser-target interaction \cite{Ott72,Sgattoni2015}, but it is not yet clear if, in general, this is the dominant process for these type of targets. Namely, the role of streaming instabilities, which are well-known to affect the electrons accelerated by the laser \cite{Gremillet1999,Silva2002,Wei2004,Califano2006, Karmakar2009,Fiuza2012,Ruyer2015}, has not yet been considered in detail for protons.

In this Letter, we report on experimental results from laser-plasma interactions using $\mu$m-scale solid-density hydrogen jets, which show that in the presence of a rear-side preplasma, the accelerated proton beams are spatially modulated by filamentary electromagnetic fields. Analytical estimates for the growth of the Weibel or current-filamentation instability \cite{Weibel1959,Fried1959,Davidson1972} (hereafter referred to as Weibel-type instabilities, WI) show that near the rear-side critical density region the fields significantly exceed those in the bulk of the target and modulate TNSA protons. Two- (2D) and three-dimensional (3D) particle-in-cell (PIC) simulations of the experimental conditions confirm our model and show good agreement with both analytical estimates and measured proton data. Simulation results further indicate that this process is relatively independent of the target material and is mostly controlled by the laser parameters and the scale of the rear-side preplasma. In particular, it will become predominant and strongly affect the quality of proton beams for scale lengths larger $\gtrsim 0.13 \lambda_0 \sqrt{a_0}$ (where $a_0$ and $\lambda_0$ are the normalized vector potential and wavelength of the laser).

\begin{figure}
\includegraphics[width=0.48\textwidth]{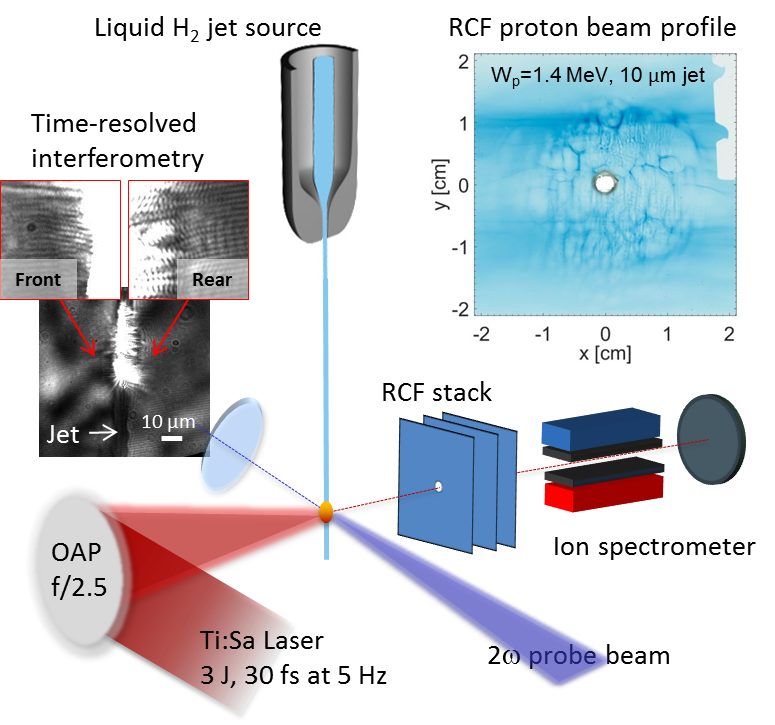}
\caption{Experimental setup: A 150-TW laser, focused onto a solid-density hydrogen jet target, accelerates protons to multi-MeV energies, measured with Thomson parabola spectrometers (TPS). The proton beam profile in the laser-forward direction is detected with RCF stacks showing strong transverse modulations. Time-resolved interferometry measurements reveal the existence of a preplasma at the time of the laser-plasma interaction.}
\label{fig:fig1}
\end{figure}

\begin{figure}
\centering
 \includegraphics[width=0.45\textwidth]{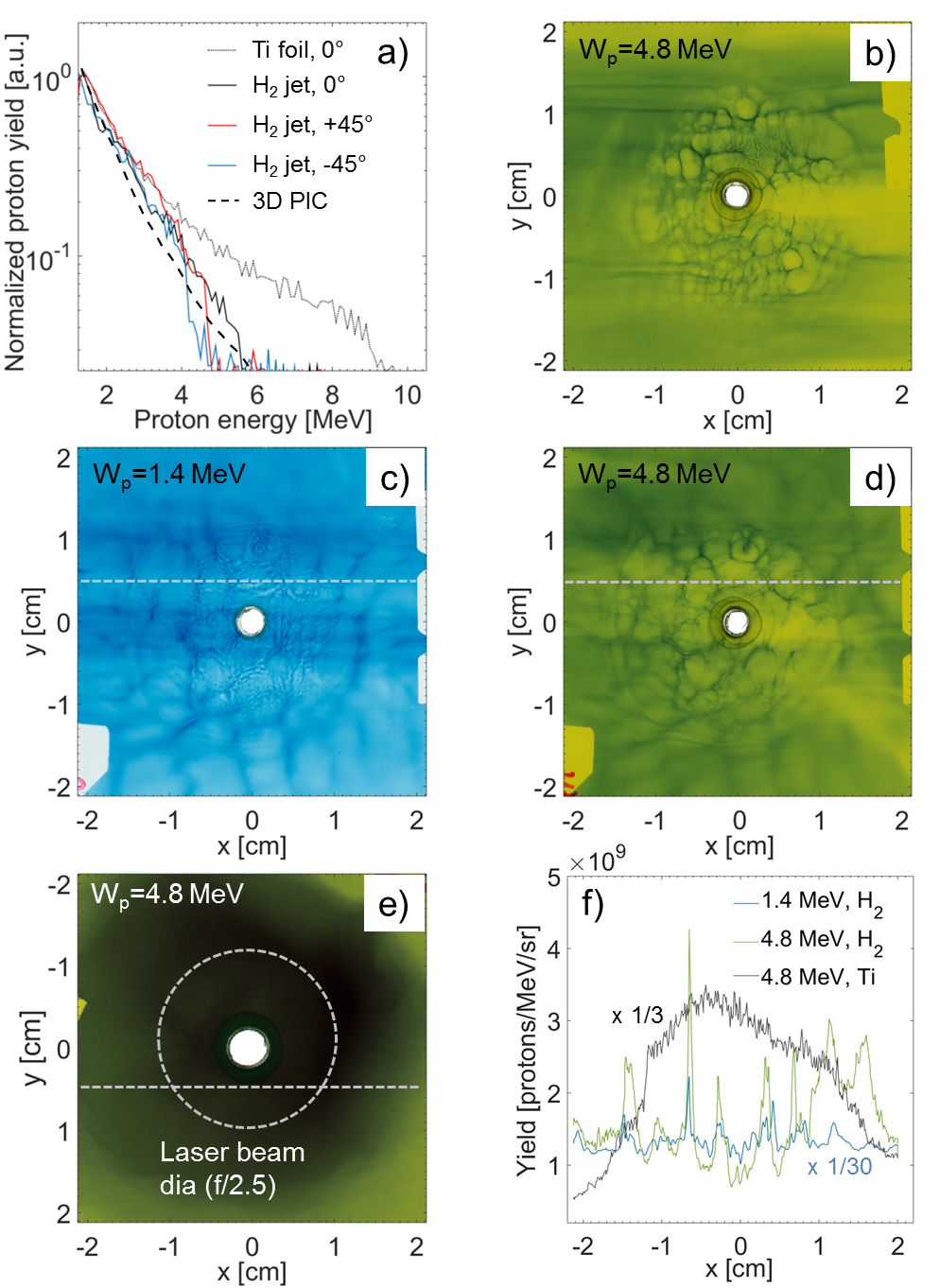}
\caption{a) Proton spectra from a 10 $\mu$m hydrogen jet, measured at different angles ($0^\circ$,$45^\circ$,$-45^\circ$) from the laser forward direction, from a 2 $\mu$m Ti foil at $0^\circ$, and from 3D PIC simulations of the hydrogen jet. b)-e) RCF stack shows the proton beam profile in laser-forward direction for b) 4.8\,MeV and 10 $\mu$m jet, c) 1.4\,MeV and 5 $\mu$m jet, d) 4.8\,MeV and 5 $\mu$m jet, and e) 4.8\,MeV and 2 $\mu$m Ti foil. f) Lineouts of the proton profiles taken at the position of the white dashed lines in c)-e).}
\label{fig:fig2}
\end{figure}

The experiment was performed using the 150-TW laser system DRACO at the HZDR. The 30 fs laser pulses of the Ti:Sapphire laser system (800\,nm) were focused with an $f/2.5$ off-axis parabolic mirror to a spot size of 3 $\mu$m (FWHM) that contains 85 \% of the pulse energy (Figure \ref{fig:fig1}). For a pulse energy of 3 J on target, the resulting average intensity is $5\times 10^{20}$ W/cm$^2$ and corresponds to a peak normalized vector potential $a_0= e E/m \omega_0 c \sim 21$. The main target consists of a solid-density hydrogen jet with a diameter of either 5 $\mu$m or 10 $\mu$m. The characterization and working principle of the cryogenic hydrogen jet source is described in Ref.\,\cite{Kim2016}. The temporal pulse contrast of the laser was measured using a third-order autocorrelator with high dynamic range. Ionization sets in at -7 ps, corresponding to a prepulse intensity of $10^{12} - 10^{13}$ W/cm$^2$. A Nomarski-interferometer setup \cite{Nomarski} using an optical probe pulse (400\,nm) is used to characterize the preplasma at $12 \mu$m from the initial target surface (see inset in Fig. \ref{fig:fig1}). The measurements reveal a significant preplasma both on front and rear sides with densities of $2.6 \times  10^{20}$ cm$^{-3}$ and $1.5 \times 10^{20}$ cm$^{-3}$, respectively  (see supplemental material). Assuming a single exponential radial profile from the edge of the cylindrical jet we estimate a plasma scale length $L_p \simeq 2.3 \mu$m on the front side and $L_p \simeq 2.1 \mu$m on the rear side, suggesting a cylindrically symmetric preplasma profile. For this preplasma level, the bulk density remains relativistically overcritical for our laser parameters. The accelerated protons were characterized using three Thomson parabola spectrometers (TPS) equipped with multi-channel plates (MCP) for fast detection, set up in the laser-forward direction (0$^\circ$) and in $\pm 45^\circ$. Proton spectra were recorded simultaneously in all TPS with a repetition rate of up to 1 Hz \cite{Gauthier2016}. All proton spectra show a similar exponential profile for all angles (0$^\circ$ and $\pm 45^\circ$), typical of TNSA, where protons are accelerated along the normal of the target surface. In the laser-forward direction, the energy cutoff is at $\sim 6$ MeV while at $\pm 45^\circ$ it is at $\sim 4$ MeV (Figure \ref{fig:fig2}a). This anisotropy of maximum proton energy is consistent with a sheath set up by the divergent fast electron stream produced by the laser \cite{Zeil2012}.

The proton beam profile was measured on a single shot basis with radiochromic film (RCF) stacks. They were installed in laser-forward direction, 55\,mm away from the interaction point covering an angle of $\pm 20^\circ$. The beam profiles for the 5 $\mu$ and 10 $\mu$m jets are shown in the inset of Figure \ref{fig:fig1} and in Figure \ref{fig:fig2}b-d. In both cases a filamentary net-like pattern with high modulation contrast can be observed, with a clear overlap of the main structures for different energy ranges (\emph{e.g.} Figure \ref{fig:fig2}c,d) and was reproduced for several shots. The size of the individual net structures on the RCF ranges between 1 and 10 mm, with the most pronounced features having $3-5$ mm (Figure \ref{fig:fig2}f).

The consistency of the accelerated proton spectrum and angular dependence with TNSA, suggests that the proton modulations are introduced on the rear side of the target, where a significant preplasma is measured. In order to investigate the importance of the rear-side plasma profile, a 2 $\mu$m thin titanium foil was irradiated under the same laser conditions. Due to the different target geometry and higher mass of Ti ions, the plasma expansion on the rear side is strongly suppressed. The measured proton spectrum has a similar shape to that obtained with the hydrogen target in the MeV energy range but extends to higher energies with a cutoff at 9 MeV due to the sharper density profile, and consistent with previous experiments \cite{Schreiber2016}. The measured protons have a smooth TNSA-type beam profile (Figure \ref{fig:fig2}e), in contrast to the hydrogen target.


By noting that the sharp modulations observed in the RCF for hydrogen correspond to caustics, which are caused by proton deflection in strong electromagnetic fields \cite{Kugland2012}, it is intuitive to consider that electron streaming instabilities may develop on the rear-side preplasma and affect the propagation of TNSA protons. In particular, WI are known to produce strong magnetic fields that affect the propagation of electrons in solid-density plasmas \cite{Gremillet1999,Silva2002,Wei2004,Califano2006,Karmakar2009}. The laser-accelerated relativistic electrons carry a current $j_f \simeq n_f c$ (where $n_f$ is the fast electron density) which is balanced by a return current sustained by the background electrons, $j_r = -j_f$. In this configuration, the cold return-current is unstable and amplifies magnetic fluctuations of the wavevector transverse to the direction of the electron beam propagation \cite{Fiuza2012,Ruyer2015}. The growth rate of the instability is $\Gamma_W \simeq (v_r/c) ~\omega_r/\sqrt{\gamma_r}$, where $\omega_r$, $v_r$, and $\gamma_r$ are the plasma frequency, mean velocity, and relativistic Lorentz factor of the return current electrons, respectively. For high-intensity lasers, $a_0 \geq 1$, the fast electron density is of the order of the critical plasma density, $n_c$. Hence, we can approximate the growth rate as $\Gamma_W \simeq \sqrt{n_c/n_r}~ \omega_0/\sqrt{\gamma_r}$, where $n_r$ is the density of the return current electrons and $\omega_0$ is the laser frequency. This scaling is only valid while the background electrons can balance the fast-electron current, i.e. when $n_r \ge n_c$. Thus, if a preplasma is present on the rear side, the growth rate is maximized at $n_r \simeq n_c$, where $v_r = - v_f \sim -c$ and $\gamma_r \sim a_0/\sqrt{2}$ \cite{Wilks2001}. Strong fields will be produced if the crossing time of relativistic return current electrons inside the rear-side preplasma exceeds the growth time of the instability. This means that the preplasma scale length needs to be larger than the critical value $L_{p,c} = c \Gamma_W^{-1} \simeq 0.13 \lambda_0 \sqrt{a_0} \simeq 0.5 \mu$m. In this case, the B-field can grow and its amplitude at saturation can be calculated by equating the Larmor radius of the electrons to the filament wavelength, $\lambda_W = 2 \pi \sqrt{\gamma_r} c/\omega_r$ \cite{Davidson1972}. This yields $B_{sat} = 1/(2\pi) \sqrt{\gamma_r n_c/n_r} m_e c \omega_0/e$, which again is maximized at $n_r \simeq n_c$. For our conditions, we have $\lambda_W \sim 3 \, \mu$m and $B_{sat} \sim 83$\,MG. The length of the filaments can be approximately calculated for an exponential density profile by considering the region where the field drops by $1/e$ of the peak value, which is $L_W \sim 2 L_p \sim 4 \, \mu$m. It is important to note that in this relativistic regime the transverse electric field associated with WI filaments can reach a strength comparable to the magnetic field and thus contribute to the proton deflection. This electrostatic field is associated with the charge imbalance produced by the magnetic pressure in the filaments \cite{dieckman2009}. Based on the balance between the transverse electric and magnetic forces on the electrons, we can estimate $E_{sat} \simeq e \lambda_W B_{sat}^2/(4 \pi m_e c^2)$ \cite{Ruyer2015}, which for our conditions yields $E_{sat} \simeq 0.5 \, \mathrm{MV}/\mu$m.

This analysis shows the critical importance of the rear-side preplasma, which allows the fields to reach amplitudes exceeding those inside the target by more than an order of magnitude and affecting the propagation of TNSA protons. The wavelength of the filamentary field structure can be inferred directly from the experimental data by assuming a virtual proton point source located near the front surface of the target, where fast electrons are produced. For a 10 $\mu$m jet, this corresponds to a distance from the virtual point source to the critical density region $\ell \sim 16.6 \,\mu$m. For a typical filament wavelength in the RCF of 5\,mm, this leads to $\lambda_W \sim 1.5\,\mu$m at $n_c$, consistent with our estimates.

The amplitude of the filamentary fields can also be estimated from the presence of caustics in the proton profile at the RCF. This means that protons from different positions in the object plane (filamentary field region) end up in the same location in the image plane (RCF). For such nonlinear features to be formed, the field amplitude needs to exceed the threshold for caustic formation. Following Ref.\,\cite{Kugland2012}, we estimate that the fields must exceed $B$ [MG] $ \geq 175 \, \alpha (\lambda_W/L_W) \sqrt{W_p[\mathrm{MeV}]}/\ell[\mu\mathrm{m}]$ or $E$ [MV/$\mu$m] $ \geq 1.9 \, \alpha (\lambda_W/L_W)^2 W_p[\mathrm{MeV}]/\ell[\mu\mathrm{m}]$, where $W_p$ is the proton energy and $\alpha = 1 ~(2.24)$ for a focusing (defocusing) field. Given that caustics are observed for proton energies up to 4.8\,MeV, this yields a minimum B-field of 9-20\,MG and E-field of 0.08-0.18 MV/$\mu$m. In fact, detailed analysis of the caustic pattern as a function of radial position indicates that the fields at the center should considerably exceed this minimum value (see supplemental material), in good agreement with the saturation values predicted theoretically.

In order to confirm our model for the field structure produced in the rear-side plasma and its role in the proton modulations we have performed 2D and 3D PIC simulations of the experimental conditions with the code OSIRIS \cite{Fonseca2002}. The laser pulse is modeled using a Gaussian-shaped profile with spot size $W_{\rm FWHM} =3 \mu$m and pulse duration of 30 fs. The FWHM laser intensity is $4\times10^{20}$ W/cm$^2$. The hydrogen target has a $5 \mu$m diameter with constant plasma density of $30 n_c$, followed by a preplasma with radial exponential profile with $L_p = 1 \, \mu$m. The 3D (2D) simulations have a box size of $95 \times 18 \times 18 \mu\mathrm{m}^3$ ($130 \times 65 \mu\mathrm{m}^2$), with a total of $4096 \times 768 \times 768$ ($8192 \times 4096$) cells, and use 8 (100) particles/cell per species.

\begin{figure}
\centering
\includegraphics[width=0.5\textwidth]{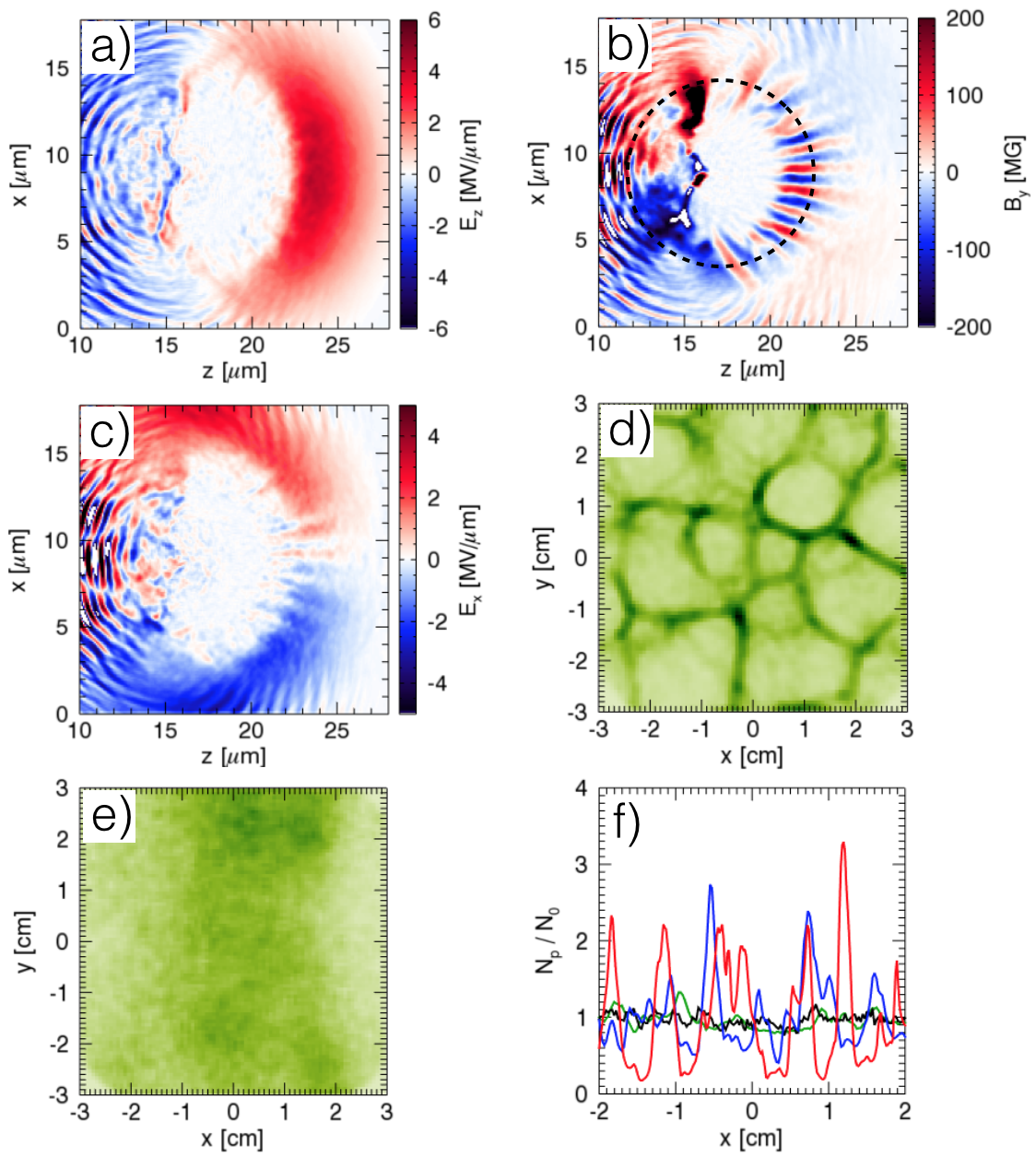}
\caption{3D PIC simulation of the interaction of a 30 fs, $4\times 10^{20}$ Wcm$^{-2}$ laser with a $30\,n_c$ hydrogen target with 5 $\mu$m diameter and a preplasma with a plasma scale length of 1 $\mu$m. The laser irradiates the target from the left side. Central slices of the a) longitudinal electric field, b) transverse magnetic field, and c) transverse electric field are shown after 68 fs of the beginning of the laser interaction. The dashed circle in b) shows the location of the critical density surface at $t = 0$. d) Projection of 4.8 MeV protons to a virtual detector placed 55 mm from the center of the target. e) Same projection as in d) but containing only protons originating from the front side of the target. f) Proton projections from 2D simulations for the same laser and plasma conditions but with variable preplasma scale length: 0 $\mu$m (black), 0.2 $\mu$m (green), 1 $\mu$m (blue), and 10 $\mu$m (red).}
\label{fig:fig3}
\end{figure}

The field structure obtained 68\,fs after the beginning of the laser-plasma interaction is shown in Figures \ref{fig:fig3}a-c. The typical TNSA E-field develops on the rear side and accelerates protons with a spectrum consistent with our experimental measurements (Figures \ref{fig:fig2}a). The transverse fields show the development of strong filamentary E- and B-fields associated with WI in the rear-side preplasma. The filaments are confined to the near-critical density region, have a length $L_W \sim 2.5 \, \mu$m, a wavelength $\lambda_W \sim 1.5 \, \mu$m, and peak amplitudes $B_{sat} \sim 110$ MG and $E_{sat} \sim 0.4 \,$MV/$\mu$m, in good agreement with our analytical estimates and experimental results.

The accelerated protons in the simulation were projected into a virtual detector placed 55 mm away from the jet axis. Figure \ref{fig:fig3}d shows the simulated proton beam profile in the detector plane. The similarity with the experimental data is striking, showing a well defined net-like structure with individual filaments ranging between 2 and 20 mm. To guarantee that the modulations are indeed produced on the rear-side plasma, we have projected separately only the protons accelerated at the front of the target, before they reach the rear side (Figure \ref{fig:fig3}e). The profile obtained is smooth, confirming that modulations are not associated with the RT instability of the front surface. Furthermore, to confirm the dominant role of the WI and that proton modulations are not affected by the choice of target material or possible ionization front instabilities, we have performed additional simulations using a plastic target (CH) with $L_p = 1 \mu$m, including self-consistently ionization and collisional dynamics. The simulations show that the ionization front is stable and that the field structure and proton spatial modulations observed are similar to the H case and indeed associated with the WI (see supplemental material). This is understood from the fact that in the presence of a rear-side preplasma the maximum growth rate and saturation amplitude of WI are mostly determined by the laser $a_0$ and $\omega_0$ (plasma critical density), not by the details of the target composition. Additionally, the ionization front instability is a relatively slow process, with typical growth time of 100 fs, and thus is not expected to be important for target thicknesses $< 30 \mu$m \cite{manclossi}.

Finally, we have performed a series of 2D PIC simulations varying the preplasma scale length between $0-10\,\mu$m to investigate its influence on the field generation and proton spatial modulations. The projected proton profiles are shown in Figure \ref{fig:fig3}f and associated B-field structures are shown in the supplemental material. For $L_p \leq 0.2 \mu$m, the profile remains smooth and no filamentary fields are produced, in agreement with the results obtained for the Ti foil. Whereas, for $L_p \geq 1 \mu$m we observe strong fields and strong spatial modulations of the proton profile, confirming our analytical prediction that the WI fields are dominant for $L_p > L_{p,c} \simeq 0.5 \mu$m.

In conclusion, we have shown that if a $\mu$m-scale rear-side preplasma is present in the interaction of intense lasers with thin solid density targets, strong filamentary fields will develop and spatially modulate accelerated proton beams. These fields are produced via the relativistic WI and are maximized near the critical density. Our results show that this process is quite general and must be taken into account when using thin targets in order to optimize the proton beam quality for multi-purpose applications. In particular, future experiments need to guarantee that the preplasma scale-length is $< 0.13 \lambda_0 \sqrt{a_0}$. Depending on the exact details of the laser prepulse and target geometry, the preplasma can be mitigated through the use of a plasma mirror and/or increased target thickness. On the other hand, we note that these results also open the possibility to study in detail the dynamics of relativistic WI by controlling the rear-side preplasma and characterizing the filamentary field structure using proton radiography.

\section*{Acknowledgments}
S. G\"ode and C. R\"odel contributed equally to this work. This work was supported by the U.S. Department of Energy SLAC Contract No. DE- AC02-76SF00515, by the U.S. DOE Office of Science, Fusion Energy Sciences under FWP 100182, FWP 100237, and ACE HEDLP Diagnostics, and by the SLAC LDRD program. It was also partially supported by EC H2020 LASERLAB-EUROPE/LEPP (contract 654148). C. R\"odel acknowledges support from the Volkswagen Foundation. The authors acknowledge the OSIRIS Consortium, consisting of UCLA and IST (Portugal) for the use of the OSIRIS 3.0 framework and the visXD framework. Simulations were conducted on Mira (ALCF supported under Contract No. DE-AC02-06CH1135) through an INCITE award and on NERSC.

\end{document}